\documentclass[12pt]{article}
\usepackage{epsfig, amsmath, amsfonts, verbatim, setspace, tabularx,multirow}
\usepackage[authoryear,round]{natbib}
\usepackage[table]{xcolor}
\usepackage{mathtools}
\usepackage{setspace}
\usepackage{lscape}
\newtheorem{assumption}{Assumption}
\newtheorem{remark}{Remark}
\newtheorem{result}{Result}
\usepackage{url}
\usepackage{graphicx} % Required for inserting images

\allowdisplaybreaks

\title{Active-Controlled Trial Design for HIV Prevention Trials with a Counterfactual Placebo}
\author{Fei Gao \and Holly Janes \and Susan Buchbinder \and Deborah Donnell}

\newcommand{\hbeta}{{\widehat\beta}}
\newcommand{\hlambda}{{\widehat\lambda}}
\newcommand{\tlambda}{{\widetilde\lambda}}
\newcommand{\hsigma}{{\widehat\sigma}}
\newcommand{\hOmega}{{\widehat\Omega}}
\newcommand{\tsigma}{{\widetilde\sigma}}
\newcommand{\tK}{{\widetilde K}}
\newcommand{\hV}{{\widehat V}}

\newcommand\consq{\stackrel{\mathclap{\normalfont\mbox{C}}}{=}}

\providecommand{\keywords}[1]
{  \small	
  \textbf{\textit{Keywords:}} #1
}

\begin{document}
\maketitle{}

\begin{abstract}
In the quest for enhanced HIV prevention methods, the advent of antiretroviral drugs as pre-exposure prophylaxis (PrEP) has marked a significant stride forward. 
However, the ethical challenges in conducting placebo-controlled trials for new PrEP agents against a backdrop of highly effective existing PrEP options necessitates innovative approaches. 
This manuscript delves into the design and implementation of active-controlled trials that incorporate a counterfactual placebo estimate — a theoretical estimate of what HIV incidence would have been without effective prevention. 
We introduce a novel statistical framework for regulatory approval of new PrEP agents, predicated on the assumption of an available and consistent counterfactual placebo estimate. 
Our approach aims to assess the absolute efficacy (i.e., against placebo) of the new PrEP agent relative to the absolute efficacy of the active control.     
We propose a two-step procedure for hypothesis testing and further develop an approach that addresses potential biases inherent in non-randomized comparison to counterfactual placebos. 
By exploring different scenarios with moderately and highly effective active controls and counterfactual placebo estimates from various sources, we demonstrate how our design can significantly reduce sample sizes compared to traditional non-inferiority trials and offer a robust framework for evaluating new PrEP agents. 
This work contributes to the methodological repertoire for HIV prevention trials and underscores the importance of adaptability in the face of ethical and practical challenges.
\end{abstract}


\section{Introduction}
Recent studies have demonstrated the high effectiveness of antiretroviral drugs (ARVs) for HIV prevention when utilized as pre-exposure prophylaxis (PrEP) \citep{grant2010preexposure, baeten2012antiretroviral, molina2015demand, mayer2020emtricitabine, landovitz2021cabotegravir, moretlwe2021long}.
Despite these successes, there remains an ongoing need for new PrEP agents that offer choice of product characteristics, longer-lasting protection, and/or fewer side effects.
Moreover, given the complexity of HIV transmission dynamics, diverse prevention strategies are essential to address the varied needs of at-risk populations \citep{bekker2022hiv}. 
Reliance on a single PrEP modality is insufficient, as no single approach can be produced and distributed at the scale necessary to meet the 95-95-95 UNAID targets \citep{roberts2022impact, mitchell2023estimating, tollett2024dynamics}.
However, designing clinical trials to evaluate new PrEP agents once we have approval of highly efficacious prevention agents poses challenges.
The gold standard randomized placebo-controlled trial randomizes individuals living without HIV to either the new PrEP agent or placebo, with prospective assessment of incident HIV based on regular HIV testing.
Placebo-controlled trials are generally deemed ethically untenable for testing new agents of a similar modality as approved agents, unless the approved agents are unavailable or undesirable for the trial population \citep{joint2012ethical}.
Instead, active-controlled trials are used, typically to establish non-inferiority; these randomize individuals without HIV to either the new agent or the approved agent (i.e., active control).
Non-inferiority (NI) trials establish efficacy of the new agent through demonstrating that it is not unacceptably less efficacious than the approved agent.
 When the approved agent is highly effective, NI trials may be prohibitively large \citep{donnell2019current, janes2023control, prudden2023perspectives}.

When the inclusion of a placebo control group is unethical, the US Food and Drug Administration has provided draft guidance for the use of external controls \citep{FDA2023externalcontrol} in establishing efficacy of a new biologic.
The primary estimand of interest for such external controls is the summary measure of clinical endpoints that would have occurred in a placebo arm of the trial, known as a ``counterfactual placebo''.
In the context of HIV prevention, various approaches to generating a counterfactual placebo HIV incidence estimate have been proposed.
For example, \cite{parkin2023facilitating} described a counterfactual placebo HIV incidence estimate obtained through a cross-sectional recency assay.
In such a trial, potential trial participants, i.e., screenees,  undergo HIV testing at screening: those who test positive for HIV are linked to HIV care services and assessed for recent infection leading to an estimate of counterfactual placebo incidence in the screenees.
Individuals without HIV are enrolled and followed prospectively while taking the new PrEP agent.
Such approach has been employed in the recent Gilead phase 3 PURPOSE1 \citep{PURPOSE1} and PURPOSE2 \citep{PURPOSE2} trials to evaluate the efficacy of a twice-yearly injectable HIV-1 capsid inhibitor, Lenacapavir.
Other promising approaches under active investigation include inferring a counterfactual placebo HIV incidence using registrational cohort data \citep{prepvacc}, external trial placebo arm data \citep{donnell2023counterfactual}, biomarker of HIV exposure \citep{mullick2020correlation, zhu2024estimating}, and the adherence-efficacy relationship of a PrEP agent \citep{glidden2021using}.
The validity and consistency of a counterfactual placebo estimate depends on the target population, the standard of care in the target population, and the reliability of historical or concurrent data on HIV incidence absent effective prevention; these issues have been discussed extensively in the literature \citep{dunn2018averted, donnell2023study}.
However, a regulatory and statistical framework for designing, analyzing, and interpreting evidence from an active-controlled trial augmented with a counterfactual placebo estimate has not yet been developed.

Even when a counterfactual placebo incidence measurement is planned to assess efficacy of a new PrEP agent, we contend that it remains crucial for the prospective trial to randomize participants to either an active control or the new agent to maintain the objective scientific integrity intrinsic to a randomized comparison.
For example, in the Gilead PURPOSE2 trial \citep{PURPOSE2}, although the primary comparison is between the incidence for participants receiving the experimental agent Lenacapavir and the counterfactual placebo incidence, participants were also randomized to receive the active control, daily oral tenofovir disoproxil fumarate/emtricitabine (TDF/FTC).
It is well-established that with a proven agent it is not ethically or clinically appropriate to deprive patients of access to the proven agent in a trial testing a new agent; an accepted practice for evaluating a new PrEP agent is a randomized trial with randomization to new versus proven agent as an active control \citep{fleming2011some}.
In addition, evaluation of efficacy by directly comparing the new agent and the counterfactual placebo without randomization carries the risk of confounding, thus leaving residual uncertainty about the validity of this comparison.
For example, as described in \cite{parkin2023facilitating}, any placebo incidence estimate derived from external data relies on assumptions regarding population similarity, adequacy of available covariates for adjustment (e.g., factors influencing HIV risk and those potentially modifying the PrEP agent's effect), and the constancy of background HIV risk across groups.
Inclusion of a randomized, active control arm and incorporation of the randomized active control incidence in evaluating the efficacy of the new agent provides some robustness against confounding compared to the direct comparison with the counterfactual placebo, a fact which will be demonstrated later in the manuscript.
Importantly, this approach also allows for a direct comparison of safety profiles between the active control and the new agent.

In this manuscript, we introduce a statistical framework that could be used as the basis for regulatory approval for the new PrEP agent in an active-controlled trial augmented with a counterfactual placebo (AC-CF).
Rather than delving into the specifics of obtaining a counterfactual placebo estimate for a particular study population, we focus on designing a trial assuming the availability and validity of such a counterfactual placebo estimate. 
Motivated by the well-established approach of assessing NI of a new intervention relative to an active control, often defined as having preserved a high fraction of the active control's effect \citep{fleming2008current,fleming2011some,FDA2016guidance}, we target the estimand of the relative absolute efficacy of the new PrEP agent relative to the active control, or relative absolute efficacy (RAE).
The RAE measures the portion of the active control's efficacy preserved by the new intervention, with the null hypothesis stating what efficacy of the new agent is not sufficient compared to the active control, a decision usually guided by clinical judgement.
We propose a two-step procedure for  evaluating this estimand in the AC-CF trial, and present the sample size calculation.
Moreover, recognizing that the counterfactual placebo is not obtained from a randomized placebo arm within the trial, we further develop an approach that is robust to bias in the counterfactual placebo estimate.
We also illustrate the proposed approaches across various realistic scenarios, highlighting the reduction in sample size compared to traditional NI trial designs.

\section{Method}\label{sec:methods}
\subsection{Non-Inferiority trial design and relative absolute efficacy hypothesis}\label{sec:ni}
\subsubsection{Non-Inferiority and relative absolute efficacy hypotheses}
We begin by providing background on the NI design which serves as both a motivation for our methods, and an important comparator design. 
NI trials are designed to determine whether a new intervention (E) is as effective as, or at least not much worse than, a standard intervention (active control, A), whose efficacy (against placebo) has been demonstrated in historical placebo-controlled trials.
They are commonly applied when randomization to placebo arm is not possible, such that absolute efficacy of E cannot be directly assessed through a placebo-controlled trial. 

To simplify the presentation of ideas we will focus on the endpoint most relevant to HIV prevention: incident HIV infection. The design (e.g., sample size) is then determined through a comparison of HIV incidences in the E and A arms, such that we evaluate the NI hypothesis:
\begin{equation}
    H_0: \log\lambda_E - \log\lambda_A\ge \delta \text{ vs. } H_a: \log\lambda_E - \log\lambda_A < \delta, \label{equ:NI}
\end{equation}
where $\lambda_A$ and $\lambda_E$ denote the incidence rates in the A and E arms in the population where the NI trial participants are recruited, respectively, and $\delta$ is an NI margin that defines the acceptable efficacy loss of E compared to A.
A common approach is to set $\delta$ to some clinically relevant portion of the effect of the active control relative to placebo, i.e., $\delta = (1-\gamma)M_1$ \citep{FDA2016guidance}.
The parameter $M_1$ is the effect of the active control relative to placebo.
The parameter $\gamma$, which is also called preservation fraction, determines the proportion of absolute efficacy of the active control that is preserved by the new intervention.
Such a parameter quantifies the largest loss of effect for the new intervention that would be clinically acceptable.

When $M_1$ is known, it can be shown that the NI hypothesis is equivalent to another hypothesis of interest given by the following Result.
\begin{result} [NI and RAE Hypotheses]\label{res:RAE}
    When the efficacy of the active control is known, i.e., $M_1 = \log\lambda_P - \log\lambda_A$, the NI hypothesis (\ref{equ:NI}) with margin $\delta = (1-\gamma)M_1$ is equivalent to the hypothesis
    \begin{equation}
    K_0: RAE\le \gamma \text{ vs. } K_a: RAE > \gamma,\label{equ:RAE}
\end{equation}
where
\[RAE = \frac{\log\lambda_P-\log\lambda_E}{\log\lambda_P-\log\lambda_A}\]
quantifies the {\it relative absolute efficacy} (RAE), i.e., the absolute efficacy of the new intervention {\it relative to absolute efficacy of the active control}.
We will call this hypothesis RAE hypothesis in the following.
\end{result}

\begin{remark}\label{rm:AER}
The RAE is closely related to the averted infections ratio \citep{dunn2018averted, dunn2019connection, glidden2020bayesian, dunn2021confidence, dunn2023interpretation}, which is defined as 
\[AIR = \frac{\lambda_P - \lambda_E}{\lambda_P - \lambda_A}.\]
That is, the RAE is the ratio of differences in log incidence, and the AIR is the ratio of differences in incidence.
\end{remark}

Result \ref{res:RAE} suggests that the RAE hypothesis is equivalent to the hypothesis (\ref{equ:NI}) evaluated in the NI trial when the efficacy of the active control is known. 
In the next section, we will show that the RAE hypothesis can also be evaluated using an NI design with an estimated NI margin under an additional assumption.

\subsubsection{``95\%-95\%'' margin and relative absolute efficacy hypothesis}
In practice, the true effect of the active control relative to placebo, i.e., $\log\lambda_P - \log\lambda_A$, is unknown.
Thus, the margin $\delta$ is usually determined based on information on efficacy of A from historical placebo-controlled trial results.
The most commonly employed ``95\%-95\%'' approach \citep{fleming2008current, fleming2011some, FDA2016guidance} chooses an NI margin 
\[\delta = (1-\gamma)(\log\tlambda_{P0}-\log\tlambda_{A0} + z_{0.025} \tsigma_{PA0}),\]
where $\tlambda_{A0}$ and $\tlambda_{P0}$ are the observed incidences in $A$ and $P$ arms, respectively, in the historical placebo-controlled trial, $\tsigma_{PA0}$ is the estimated standard error for $\log\tlambda_{P0}-\log\tlambda_{A0}$, and $z_\alpha$ denote the $\alpha$-quantile of the standard normal distribution.

After the NI margin is chosen, for example, based on the ``95\%-95\%'' approach, the hypothesis (\ref{equ:NI}) can be assessed through the test statistic
\[T_{NI} = \frac{\log\hlambda_E - \log\hlambda_A - \delta}{\hsigma_{EA}},\]
where $\hlambda_E$ and $\hlambda_A$ denote the observed incidence of $E$ and $A$ arms, respectively, in the NI trial, and $\hsigma_{EA}$ is the estimated standard error of $\log\hlambda_E - \log\hlambda_A$.
Specifically, one would reject the null hypothesis (\ref{equ:NI}) if $\{T_{NI}\le z_\alpha\}$.

Note that here the NI design is based on a fixed NI margin $\delta$, such that it is {\it conditional} on the historical trial data.
That is, no matter how the margin $\delta$ is chosen, the design would have (theoretical) protected conditional type-1 error and desired conditional power for evaluating hypothesis (\ref{equ:NI}).
Under the following {\it constancy assumption}, we can then show that for an NI design using a margin $\delta$ based on the ``95\%-95\%'' method, the RAE hypothesis can be assessed through the test statistic $T_{NI}$.
The derivations are shown in Appendix \ref{append:NI_deriv}.

\begin{assumption}[Constancy Assumption]
The absolute efficacy of the active control in the historical placebo-controlled trial is the same as the absolute efficacy of the active control in the NI trial, that is,
\[\log\lambda_{P} - \log\lambda_A = \log\lambda_{P0} - \log\lambda_{A0},\]
where $\lambda_{A0}$ and $\lambda_{P0}$ are the incidences in A and P arms, respectively, in the historical placebo-controlled trial.
\end{assumption}
\begin{result}[RAE Hypothesis in NI Designs with ``95\%-95\%'' Margin]\label{res:RAE2}
    Under the constancy assumption (Assumption 1), the RAE hypothesis (\ref{equ:RAE}) can be assessed through the test statistic $T_{NI}$ in an NI trial design with margin determined through the “95\%-95\%” approach.
    Specifically, the test $\{T_{NI}\le z_\alpha\}$ preserves type-1 error in assessing the RAE hypothesis.
    It has the same power to evaluate the alternative hypothesis that matches the alternative hypothesis in (\ref{equ:NI}), i.e., when  $\gamma^*=1 - \delta^* / (\log\lambda_P -\log\lambda_A)$, where $\delta^*$ and $\gamma^*$ define the alternative hypotheses of interest for (\ref{equ:NI}) and ($\ref{equ:RAE}$), respectively.
\end{result}

The type-1 error (and power) against the RAE hypothesis is {\it unconditional}, i.e., it accounts for the variability from the historical placebo-controlled trial.
As shown analytically (in Appendix \ref{append:NI_deriv}) and numerically (in Section \ref{sec:simu}), the test is very conservative in preserving the type-1 error for the RAE hypothesis under the constancy assumption.

\subsection{Leveraging counterfactual placebo in active-controlled trial} \label{sec:cf}
Suppose that a counterfactual placebo incidence estimate is available for the population of an active-controlled trial. 
We propose considering the RAE hypothesis in such a design due to several reasons. First, evaluating the absolute efficacy of the new intervention remains the ultimate goal for guiding policy making and individual decision-making around the intervention, and the RAE hypothesis directly relates to this absolute efficacy. 
Second, as demonstrated in Section \ref{sec:ni}, the RAE hypothesis is closely connected to and evaluated in the NI design, which is a common design for efficacy evaluation of a new intervention in the presence of an effective active control. 
Using the same RAE hypothesis provides an advantage in interpreting results for practitioners familiar with the NI trial design.
Third, the RAE naturally utilizes all available incidence estimates in the active-controlled trial design with a counterfactual placebo and is more robust against the bias of the counterfactual placebo estimate compared to an absolute efficacy estimand, as will be demonstrated further in Section \ref{sec:simu_onearm}.

We propose a formal testing procedure to assess the RAE hypothesis based on a study design with randomized active control augmented with counterfactual placebo. 
Specifically, if $\log\lambda_P-\log\lambda_A>0$, hypothesis (\ref{equ:RAE}) is equivalent to the hypothesis
\[\tK_0:(1-\gamma)\log\lambda_P-\log\lambda_E+\gamma\log\lambda_A\le 0\text{ vs. }\tK_a: (1-\gamma)\log\lambda_P-\log\lambda_E+\gamma\log\lambda_A> 0.\]
This motivates the following two-step procedure.

Let $\hlambda_P$ be the counterfactual placebo incidence estimate.
In the first step, we test if 
\[H_0^{AS}: \log\lambda_P-\log\lambda_A\le0 \text{ vs. }H_a^{AS}: \log\lambda_P-\log\lambda_A>0\] based on the test statistic 
\[T_{PA} = \frac{\log\hlambda_P- \log\hlambda_A}{\hsigma_{PA}},\]
and reject the hypothesis if $T_{PA}\ge -z_\alpha$, where $\hsigma_{PA}$ is the estimated standard error of $\log\hlambda_P - \log\hlambda_A$.
This hypothesis is called ``assay sensitivity'' (AS) hypothesis in NI trial literature, since it describes the ability of a specific trial to detect a difference between treatments if one exists (that is, assay is working and can detect a difference) \citep{temple2000placebo, d2003non}.
If this hypothesis is not rejected, we accept the null hypothesis $K_0$; otherwise, we proceed to the second step.
In the second step, we consider the test statistic
\[T_{CF} = \frac{(1-\gamma)\log\hlambda_P - \log\hlambda_E + \gamma\log\hlambda_A}{\hV_\gamma^{1/2}},\]
where $\hV_\gamma$ is the estimated variance of the numerator.
We will then reject the hypothesis $K_0$ if $T_{CF}\ge -z_\alpha$.

Based on the two-step procedure, we will reject the hypothesis $K_0$ if both $T_{PA}\ge -z_\alpha$ and $T_{CF}\ge -z_\alpha$.
Note that the null hypothesis of interest (\ref{equ:RAE}) can be written as $K_0 = (H_a^{AS}\cap\tK_0)\cup (H_0^{AS}\cap\tK_a)$, which is a subset of $\tK_0\cup H_0^{AS}$, i.e., the joint set of the null hypotheses in the two steps.
Therefore, the two-step procedure preserves the overall type-1 error, which is given in the following Result.
\begin{result}[Characteristic of the Proposed Two-Step Testing Procedure]
The overall type-1 error for the RAE hypothesis using the proposed two-step procedure is bounded above by the maximum of the type-1 errors in the two steps such that the overall type-1 error is preserved at level $\alpha$. 
\end{result}
In simulation studies, we found that the empirical overall type-1 error is indeed quite close to the nominal level.

The proposed hypothesis and testing procedure are closely related to those in the ``fraction approach'' proposed to assess NI of a new treatment in a three-arm trial that includes randomization to a placebo \citep{pigeot2003assessing, koch2004hypothesis}.
In the three-arm trial, the observed incidence in the placebo arm is automatically consistent due to randomization, while for the current design the consistency of the counterfactual placebo estimate is assumed and may be violated.

\subsection{Sample size determination}
In this section, we calculate the sample size for the proposed AC-CF design.
Specifically, we consider the case when $\log\hlambda_P$, $\log\hlambda_E$, and $\log\hlambda_A$ are mutually independent, with corresponding variances 

\[\sigma_P^2 = c_{PO}/N + c_{P1},\ \sigma_E^2 = c_E/N,\text{ and } \sigma_A^2 = c_A/N,\]
where $N$ is the size of the active-controlled trial. 
This general specification of $\log\hlambda_P$, $\log\hlambda_E$, and $\log\hlambda_A$ reflects the practice for active-controlled HIV prevention trials and includes common candidates for counterfactual placebo incidence estimates.
For example, if the incidences $\hlambda_E$ and $\hlambda_A$ are estimated by the number of cases divided by the number of person-years in each arm, then $c_E = \lambda_E^{-1}/2$ and $c_A = \lambda_A^{-1}/2$.
If the counterfactual placebo estimate is based on a prospective follow-up of individuals over $n$ person-years, then $c_{P0} = 0 $ and $c_{P1} = \lambda_P^{-1}/n$.
If the counterfactual placebo estimate is based on applying cross-sectional HIV recency testing to the screening population in the trial, then we have 
\begin{align*}
    c_{P0} =& \frac{1}{p}\left\{\frac{P_R(1-P_R)}{(P_R-\beta_{T})^2} + \frac{1}{1-p} + \frac{(1-p)\sigma^2_{\hbeta_T}}{(P_R-\beta_{T})^2}\right\},\\
    c_{P1} =& \frac{\sigma_{\hOmega_T}^2}{(\Omega_T - \beta_TT)^2} + \sigma_{\hbeta_T}^2\left\{\frac{(\Omega_T - \beta_TT)^2}{(P_R-\beta_T)^2(\Omega_T-\beta_TT)^2}\right\},
\end{align*}
where $p$ is the HIV prevalence in the screening population, $P_R$, $\Omega_T$, $\beta_T$, and $T$ are parameters related to HIV recency assay that was described in \cite{gao2021sample}.

Consider the power of the proposed testing procedure against the alternative hypothesis $K_a^*: (\log\lambda_P-\log\lambda_E)/(\log\lambda_P-\log\lambda_A)= \gamma^* $.
The power of the two-step procedure satisfies
\begin{align*}
    \Pr(T_{PA}\ge -z_\alpha, T_{CF}\ge -z_\alpha | K_a^*) \ge &  1 - \Pr(T_{PA}< -z_\alpha| K_a^*) - \Pr(T_{CF}< -z_\alpha | K_a^*)\\
     = &\Pr(T_{CF} \ge -z_\alpha| K_a^*) - \Pr(T_{PA}< -z_\alpha | K_a^*).
\end{align*}
That is, if a sample size $N$ would allow $\Pr(T_{CF} \ge -z_\alpha| K_a^*) - \Pr(T_{PA}< -z_\alpha | K_a^*) \ge \beta$, it will always guarantee a $\beta$-power against $K_a^*$ based on the proposed testing procedure.
Based on the derivations in the Appendix \ref{append:AC-CF}, the sample size $N$ satisfies 
\begin{align}
    &\Phi\left(z_\alpha + \frac{(\gamma^*-\gamma)(\log\lambda_P - \log\lambda_A)}{\sqrt{\{(1-\gamma)^2c_{P0}+c_E+\gamma^2 c_A\}/N + (1-\gamma)^2c_{P1}}}\right) \nonumber\\
    &\qquad \qquad+ \Phi\left(z_\alpha + \frac{\log\lambda_P- \log\lambda_A}{\sqrt{(c_{P0} + c_A)/N + c_{P1}}}\right)= 1+\beta.\label{equ:samplesize}
\end{align}
The equation is solved numerically via a grid search for the sample size $N$ of the active-controlled trial.

\subsection{Robustness against bias of counterfactual placebo} \label{sec:AC_CF_9595}
A common criticism in utilizing a counterfactual placebo incidence estimate is that it may fail to provide a consistent estimate of the placebo incidence for the trial population, unlike the consistency guaranteed by randomization to a placebo arm.
Therefore, a study design robust to bias of the counterfactual placebo estimate is desirable.
In this section, we introduce a study design  that we term the conservative AC-CF design. 
This methodology is inspired by the "95\%-95\%" approach for selecting the NI margin.

Let $\hlambda_P^L = \hlambda_P\exp(z_{0.025} \hsigma_{P})$ be the lower bound of the 95\% confidence interval for the counterfactual placebo incidence.
We modify the procedure of testing RAE by the following two-step procedure.
In the first step, we test if 
\[H_0^{AS}: -\log\lambda_A\le - \log \hlambda_P^L\text{ vs. }H_a^{AS}: -\log\lambda_A > - \log\hlambda_P^L\] based on the test statistic 
\[T_{PA}' = \frac{\log\hlambda_P^L - \log\hlambda_A}{\hsigma_{A}},\]
and reject the hypothesis if $T_{PA}'\ge -z_\alpha$.
If this hypothesis is not rejected, we accept the null hypothesis $K_0$.
Otherwise, we proceed to the second step and consider the test statistic
\[T_{CF}' = \frac{(1-\gamma)\log\hlambda_P^L - \log\hlambda_E + \gamma\log\hlambda_A}{\hV_\gamma'^{1/2}},\]
where $\hV_\gamma'$ is the estimated variance of $\log\hlambda_E + \gamma\log\hlambda_A$, and reject the hypothesis $K_0$ if $T_{CF}'\ge -z_\alpha$.
That is, we will reject the hypothesis $K_0$ if both $T_{PA}'\ge -z_\alpha$ and $T_{CF}'\ge -z_\alpha$.

Note that the proposed testing procedure has two differences from that in Section \ref{sec:cf}.
First, the lower bound $\hlambda_P^L$ replaces the estimated $\hlambda_P$ in the two test statistics.
Second, the denominators which reflect the variability of the numerators are modified to ignore variability from the counterfactual placebo incidence estimate, i.e., $\hlambda_P^L$ is treated as a fixed constant in building the test statistics.
For example, $\hV_\gamma$ in the denominator of $T_{CF}$ is replaced by $\hV_\gamma'$ that includes the variability from $\hlambda_E$ and $\hlambda_A$ only.
The modification is similar in spirit to the ``95\%-95\%'' margin approach in the NI design, where the margin $\delta$ is based on the lower bound of the 95\% confidence interval of the risk ratio and it is treated as a fixed constant in assessing the hypothesis (\ref{equ:NI}).
It then leads to a similar conservative type-1 error in assessing the RAE hypothesis, given in the following Result. 
The technical derivations are given in Appendix \ref{append:cons_AC-CF}.
In the simulation studies, we also demonstrate the conservativeness of the design in realistic settings.

\begin{result}[Characteristic of the Modified Two-Step Testing Procedure]
The proposed modified two-step procedure has conservative type-1 error in assessing the RAE hypothesis  when the counterfactual placebo estimate is consistent.
Specifically, the analytical value of the type-1 error is always no larger than the nominal level and depends on the relative variabilities of $\hlambda_A$, $\hlambda_E$, and $\hlambda_P$.
\end{result}

Based on such testing procedure, the sample size $N$ can be determined by numerically solving
\begin{align*}
    &\Phi\left(\frac{\sqrt{(c_E+\gamma^2 c_A)/N }+ (1-\gamma)\sqrt{c_{P0}/N+c_{P1}}} {\sqrt{(c_E+\gamma^2 c_A)/N + (1-\gamma)^2(c_{P0}/N+c_{P1})}} z_\alpha + \frac{(\gamma^*-\gamma)(\log\lambda_P - \log\lambda_A)}{\sqrt{(c_E+\gamma^2 c_A)/N + (1-\gamma)^2(c_{P0}/N+c_{P1})}}\right) \nonumber\\
    &\qquad \qquad+ \Phi\left(\frac{\sqrt{c_{P0} /N + c_{P1}}+\sqrt{c_A/N}}{\sqrt{(c_{P0} + c_A)/N + c_{P1}}}z_\alpha + \frac{\log\lambda_P- \log\lambda_A}{\sqrt{(c_{P0} + c_A)/N + c_{P1}}}\right)= 1+\beta.
\end{align*}
The derivation for the sample size is given in Appendix \ref{append:cons_AC-CF}.
The resulting sample size in the active-controlled trial is always larger than that from (\ref{equ:samplesize}).

\section{Simulation Studies} \label{sec:simu}
In this section, we provide numerical studies to assess the performance of different designs in evaluating the RAE hypothesis.
We consider the proposed AC-CF and conservative AC-CF designs, and compare them with the NI trial design, the classical approach in the context of a highly-effective active control.
Specifically, we calculate the sample size and empirical type-1 error and power for the designs when corresponding assumptions hold, and evaluate the robustness of the designs when the assumptions are violated.
In addition, we compare the proposed AC-CF design with the single-arm trial design with a counterfactual placebo.
\subsection{Simulation setting}\label{sec:simu_para}
We assume a placebo HIV incidence in the trial population of $\lambda_P = 0.03$ cases/person-year (PY).
We consider a moderately efficacious active control with 55\% prevention efficacy, i.e., $\log\lambda_P -\log\lambda_A = \log(2.2)$, such that $\lambda_A = 0.014$  cases/PY.
We will evaluate the RAE hypothesis (\ref{equ:RAE}) with the null hypothesis parameter $\gamma = 0.5$, which corresponds to an efficacy of the new PrEP agent under the null hypothesis of 33\%, i.e. an efficacy considered insufficient in the context of the proven active control.
We wish to achieve $\beta$-power  under level $\alpha = 0.025$, against an alternative hypothesis with \[\frac{\log\lambda_P - \log\lambda_E}{\log\lambda_P -\log\lambda_A} = \gamma^* = 1.36.\]
The choice of $\gamma^*$ corresponds to 66\% efficacy of the new PrEP agent, or, equivalently, an NI margin of $\delta^* = 0.75$.
The simulation setting is very similar to the design of HPTN 083 study \citep{landovitz2021cabotegravir}, which compares the safety and efficacy of injectable Cabotegravir (CAB-LA) and daily oral TDF/FTC for PrEP in cisgender men and transgender women who have sex with men without HIV.
We will consider designs with $\beta$ = 0.8 and 0.9.

\subsubsection{Non-inferiority design}
To evaluate the design of a traditional NI trial, we suppose a historical placebo-controlled trial for the active control was conducted in a population with a higher placebo HIV incidence of $\lambda_{P0}= 0.05$ cases/PY.
This reflects the scenario of declining HIV incidence since the historical trial was conducted and emphasizes that the design of an NI trial does not necessitate a similar placebo HIV incidence between the historical and current trials.
We assume the constancy assumption holds such that $\log\lambda_{P0} - \log\lambda_{A0} = \log(2.2)$ and  $\lambda_{A0} = 0.023$ cases/PY.
Thus, we assume that the historical data support the design of the NI trial in that, even though the placebo incidence does not ``transport'' from the historical trial to the current NI trial setting, the active control effect does. 
The historical placebo-controlled trial is generated with a total of 3,610 PYs equally distributed between the active control and placebo arms, such that the variability of the estimated absolute efficacy of the active control is comparable to that used in the sample size determination for the HPTN 083 study \citep{landovitz2021cabotegravir}.

We consider the case where the NI margin $\delta$ is set using the ``95\%-95\%'' approach based on the estimates from the historical placebo-controlled trial and $\delta^* = 0.75$ that is consistent with the RAE alternative hypothesis.
The NI trial is designed based on the same design level and power ($\alpha = 0.025$, $\beta = 0.8$ or 0.9) for evaluating the NI hypothesis.
For each simulation, we simulate the historical placebo-controlled trial data based on which we will calculate the corresponding margin $\delta$.
Since the NI trial sample size (e.g., total PYs and total number of events (\#Events)) depends on the estimate from the historical trial and varies across simulations, we will report the average total PYs and average \#Events among simulations.

\subsubsection{Counterfactual placebo}
For the AC-CF and conservative AC-CF designs, we will consider two scenarios where the counterfactual placebo incidence estimates are generated from two different sources.
For both scenarios, the counterfactual placebo incidence is 0.03 cases/PY, such that the estimate is consistent for the placebo incidence in the active-controlled trial.
\paragraph{External follow-up approach}
The counterfactual placebo incidence estimate is based on follow-up data from an external population, e.g., from a registration cohort that is concurrent with the current trial.
We assume we collect 1,805 PYs of follow-up to estimate the counterfactual placebo incidence.
This is the same amount of follow-up as that in the placebo arm of the historical placebo-controlled trial, such that the amount of statistical information is comparable.
\paragraph{Recency testing approach}
The counterfactual placebo incidence estimate is based on recency testing at screening for the active-controlled trial. 
Every eligible screened individual with HIV is tested using the recency assay and every eligible screened individual without HIV  is enrolled in the active-controlled trial.
We suppose that the setting is that of a subtype B epidemic with 15\% HIV prevalence, such that the MDRI is 142 days and FRR is 1\% using the Limiting Antigen (LAg) Avidity (Sedia HIV-1 LAg Avidity EIA; Sedia Biosciences Corporation, Portland, OR, USA) ODn $\le$ 1.5 and viral load >1,000 copies/mL with cutoff T = 2 years \citep{grebe2019impact}.

When the counterfactual placebo estimate is based on the recency testing approach, the precision of the counterfactual placebo incidence estimate depends on the number of participants screened, which is then dependent on the duration of individual follow-up, denoted as $\tau$, given fixed total follow-up PYs.
Therefore, the sample sizes for the AC-CF and conservative AC-CF designs further depend on $\tau$ in addition to other design parameters.
We consider $\tau = $ 1 and 2 years in the simulations.

\subsubsection{Comparison of external information}
Table \ref{tab:ext_info_compare} summarizes the PYs and \#Events for participants receiving either active control or placebo in the external studies utilized for different trial designs, which are indicative of the amount of external statistical information.
In this simulation setting, the counterfactual placebo estimate based on either the external follow-up or the recency testing approach contains comparable or less statistical information compared to that from the historical placebo-controlled trial that supports the design of the NI trial.

\begin{table}[htbp]
\small
\caption{Summary of statistical information in the external studies that are used to estimate margin for the NI trial and counterfactual placebo HIV incidence for the AC-CF and conservative AC-CF trials.
For the recency testing approach for generating counterfactual placebo estimate, we display the number of recent infections as \#Events.
It is dependent on the duration of follow-up in the active-controlled trial, such that the mean and range in the simulated settings is provided.}\label{tab:ext_info_compare}
\begin{center}
\begin{tabular}{llccccc}
\hline
\multirow{2}{*}{Trial Design} & \multirow{2}{*}{External Data} & \multicolumn{2}{c}{Active Control} && \multicolumn{2}{c}{Placebo}\\\cline{3-4}\cline{6-7}
&&PY & \#Event && PY &\#Event\\\hline
NI &Historical trial for active control & 1,805 & 41 && 1,805 & 90\\\\
\multirow{2}{*}{(Conservative) AC-CF} & External follow-up data & 0 & 0&& 1,805 & 54\\
& Recency testing data at screening & 0&0&& - & 80 (43-130)\\
\hline
\end{tabular}
\end{center}
\normalsize
\end{table}

\subsection{Simulation Results}
Table \ref{tab:simu_alldesigns} shows the calculated sample size (total PYs and \#Events) for different designs, along with the empirical and analytical type-1 error and power for assessing the RAE hypothesis, estimated empirically based on 10,000 simulations.
For the AC-CF and conservative AC-CF designs with the counterfactual placebo estimate based on recency testing approach, the number of participants screened, number of participants with HIV that test recent, along with other important sample sizes in the screening phase are given in Table \ref{tab:simu_RA} in the Appendix.

\begin{landscape}
    \begin{table}[htbp]
\caption{Summary for different designs.
For all designs, total PYs and total \#Events include the prospective follow-up time and expected number of HIV infections under the alternative hypothesis, respectively, in the active-controlled trial only.
}\label{tab:simu_alldesigns}
\begin{center}
\begin{tabular}{llccccccc}
\hline
\multirow{2}{*}{Design}  &\multirow{2}{*}{Counterfactual placebo approach}& \multirow{2}{*}{Total PYs} & \multirow{2}{*}{Total \#Events} & \multicolumn{2}{c}{Empirical} &&\multicolumn{2}{c}{Analytical}\\\cline{5-6}\cline{8-9}
&& && Type-1 error & Power&& Type-1 error & Power\\\hline
&&&&\multicolumn{5}{c}{Design power = 80\%}\\\hline
NI & - & 12,016 & 255 &	0.0034	&	0.801	&&	0.0041	&	0.8\\\\
\multirow{3}{*}{AC-CF} & External follow-up & 4,942 & 105 &	0.021	&	0.844&& 0.025 & 0.8\\
 & Recency testing ($\tau = 1$ year) & 5,432 & 115 & 0.022	&	0.835 && 0.025 & 0.8\\
 & Recency testing ($\tau = 2$ year) & 6,668 & 142 &	0.021	&	0.818&& 0.025 & 0.8\\\\
\multirow{3}{1.8cm}{Conservative AC-CF}& External follow-up & 8,205 & 174 & 0.0038	&	0.822	&&	0.0028	&	0.8\\
 & Recency testing ($\tau = 1$ year) & 8,266 & 176 &	0.0031	&	0.834	&&	0.0028	&	0.8\\
 & Recency testing ($\tau = 2$ year) & 10,468 & 222 & 0.0031	&	0.825	&&	0.0031	&	0.8\\\hline
&&&&\multicolumn{5}{c}{Design power = 90\%}\\\hline
NI & - & 16,190 & 344& 0.0025	&	0.904 && 0.0036	&	0.9	 \\\\
\multirow{3}{*}{AC-CF} & External follow-up & 6,554 & 139 &	0.022	&	0.921&& 0.025 & 0.9\\
 & Recency testing ($\tau = 1$ year) & 6,868 & 146 &	0.021	&	0.921&& 0.025 & 0.9\\
 & Recency testing ($\tau = 2$ year) & 8,396 & 178 &	0.021	&	0.902&& 0.025 & 0.9\\\\
\multirow{3}{1.8cm}{Conservative AC-CF}& External follow-up &  10,938 & 232 & 0.0033	&	0.899	&&	0.0029	&	0.9\\
 & Recency testing ($\tau = 1$ year) & 10,132 & 215 & 0.0043	&	0.918	&&	0.0028	&	0.9\\
 & Recency testing ($\tau = 2$ year) & 12,780 & 272 & 0.0029	&	0.903	&&	0.0031	&	0.9\\
\hline
\end{tabular}
\end{center}
\normalsize
\end{table}
\end{landscape}

For all designs, the empirical type-1 error is close to or smaller than the nominal level and the empirical power is close to the nominal level, indicating that all designs achieve the design goal of evaluating the new PrEP agent with the given calculated sample sizes.
The analytical type-1 errors and powers match the empirical estimates.
The type-1 errors for the NI and conservative AC-CF trial designs are much smaller than the nominal level, verifying that the NI and conservative AC-CF trial designs allow for assessment of the RAE hypothesis with conservative type-1 error.

Comparing different designs, the prospective follow-up time in the AC-CF design is 45\% - 60\% smaller than that in the NI design.
The sample size from the conservative AC-CF design is larger than that from the AC-CF design based on the same counterfactual placebo estimate.
The total follow-up PYs in the conservative AC-CF design is still smaller than that of the NI design (approximately 13\% - 37\% reduction).

\subsection{Violation of assumptions}\label{sec:simu_violation}
Assessing the RAE hypothesis with an NI design relies on the validity of the constancy assumption, while assessing the same hypothesis with the two AC-CF-type designs relies on the assumption of consistency of the counterfactual placebo incidence estimate.
In this section, we evaluate the performance of the designs under violation of either assumption.
Here, we focus on the AC-CF and conservative AC-CF designs with a counterfactual placebo incidence estimated from the external follow-up approach.
As mentioned in Section \ref{sec:simu_para}, the counterfactual placebo estimate is based on the same PYs of follow-up data as accrued in the placebo arm of the historical placebo-controlled trial that supports the NI trial design.

We assume that the trials (NI, AC-CF, conservative AC-CF) are designed based on the design parameters in Section \ref{sec:simu_para}, and we target 80\% power.
The corresponding sample sizes are given in Table \ref{tab:simu_alldesigns}: 12,016, 4,942, and 8,205 prospective follow-up PYs are required for the NI, AC-CF, and conservative AC-CF designs, respectively.
We evaluate the performance of those designs when the incidences for the trial population deviate from the design parameters.
Specifically, the trials are designed based on $\lambda_P$ = 0.03 cases/PY and $\lambda_A$ = 0.014 cases/PY, and we suppose that the trial is indeed associated with different values of $(\lambda_P,\lambda_A)$ that are potentially different from the design parameters.
For each combination of $(\lambda_P,\lambda_A)$, we evaluate the empirical type-1 errors under the null hypothesis with $\lambda_E = \lambda_A^\gamma \lambda_P^{1-\gamma}$, and the empirical power under the alternative hypothesis with $\lambda_E = \lambda_A^{\gamma^*} \lambda_P^{1-\gamma^*}$, where $\gamma = 0.5$ and $\gamma^* = 1.36$ as specified in Section \ref{sec:simu_para}.

\begin{figure}[htbp] \centering
\includegraphics[width=\textwidth]{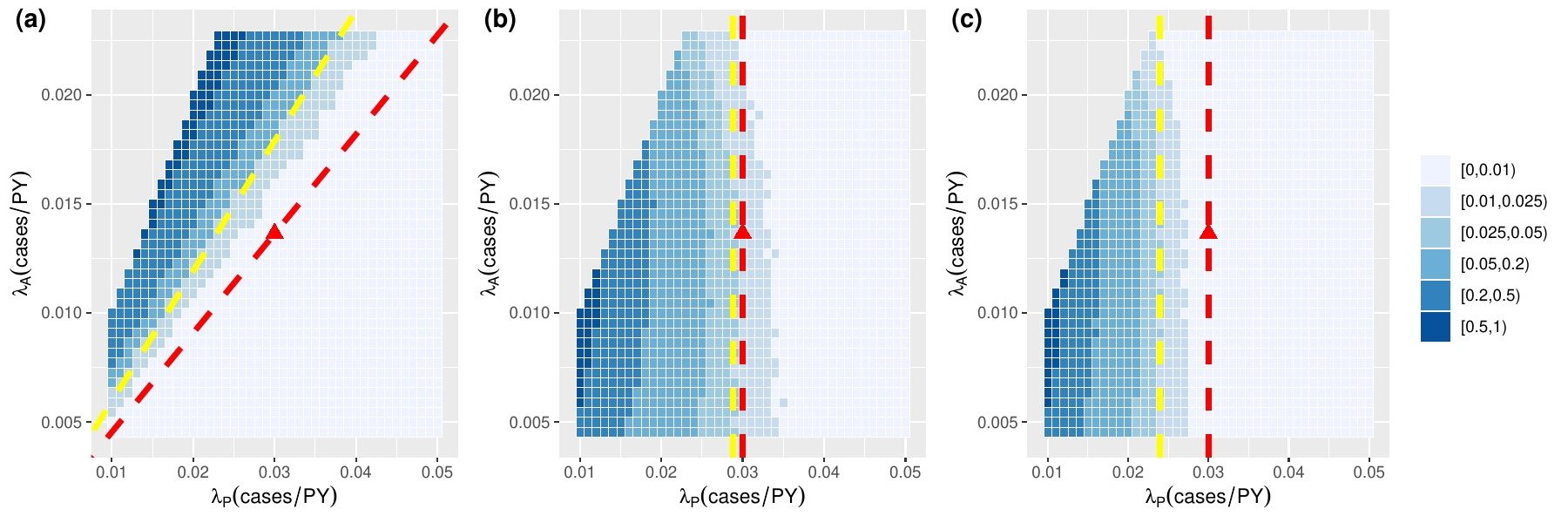}
\caption{Empirical type-1 error for assessing the RAE hypothesis (\ref{equ:RAE}) with different values of $(\lambda_P,\lambda_A)$.
(a) NI design with an average of 12,010 PYs and 95\%-95\% margin based on a historical placebo-controlled trial with 3,610 PYs. (b) AC-CF design with 4,942 trial PYs and a counterfactual placebo estimate based on external follow-up data with 1,805 PYs. (c) Conservative AC-CF design with 8,205 trial PYs and a counterfactual placebo estimate based on external follow-up data with 1,805 PYs. 
The red triangles correspond to the design parameters.
The red lines correspond to settings when corresponding assumption holds (constancy assumption for (a) NI design and consistency of counterfactual placebo estimate for (b) AC-CF and (c) conservative AC-CF designs).
The yellow lines correspond to fitted lines with the empirical type-1 error close to the nominal 0.025 level.}\label{fig:Violation}
\end{figure}

Figure \ref{fig:Violation} shows the plot of the empirical type-1 error for the three designs (NI, AC-CF, and conservative AC-CF) with different combinations of $(\lambda_P,\lambda_A)$.
Here, we consider only the scenarios in which the active control reduces incidence, i.e., $\lambda_A\le \lambda_P$.
In each sub figure, the red triangle corresponds to the design parameters; the red line corresponds to the settings when the corresponding assumption holds, i.e., constancy assumption for NI design and consistency of counterfactual placebo estimate for AC-CF-type designs; and the yellow line correspond to a fitted line with the empirical type-1 error close to the nominal level.

As expected, the type-1 errors are preserved when the corresponding assumption holds for the three designs, i.e., when the active control is as effective as assumed in the NI design, and when the counterfactual placebo incidence reflects the placebo incidence in the AC-CF and conservative AC-CF designs.
When the placebo incidence is lower than the counterfactual placebo incidence, the AC-CF design has inflated type-1 error.
The NI design and conservative AC-CF design are more robust to type-I error inflation since they allow a larger amount of assumption violation compared to the AC-CF design.
Specifically, the NI design has protected type-1 error when the active control has a prevention efficacy of no smaller than 40.4\% (the prevention efficacy for the active control in the historical placebo-controlled trial is 54.5\%), and the conservative AC-CF design has protected type-1 error when the placebo incidence is no smaller than 0.024 cases/PY (the counterfactual placebo incidence is 0.03 cases/PY).

\begin{figure}[htbp] \centering
\includegraphics[width=\textwidth]{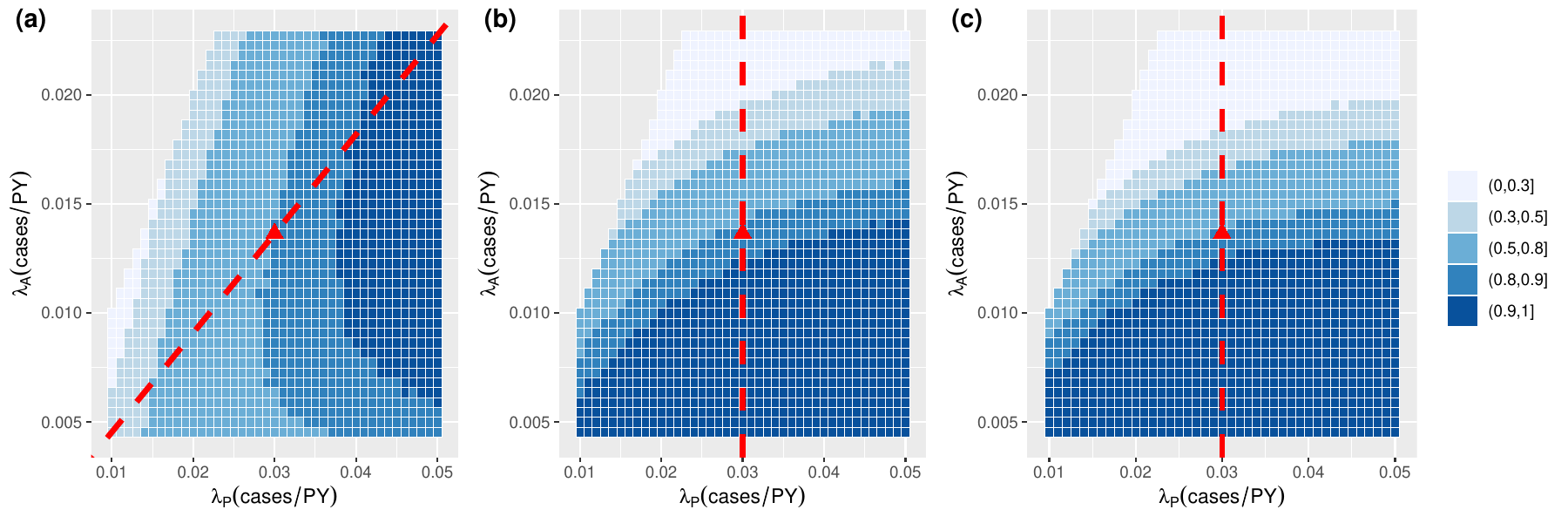}
\caption{Empirical power for assessing the RAE hypothesis (\ref{equ:RAE}) with different values of $(\lambda_P,\lambda_A)$.
(a) NI design with an average of 12,010 PYs and 95\%-95\% margin based on a historical placebo-controlled trial with 3,610 PYs. (b) AC-CF design with 4,942 trial PYs and counterfactual placebo estimate based on follow-up data with 1,805 PYs. (c) Conservative AC-CF design with 8,205 trial PYs and counterfactual placebo estimate based on follow-up data with 1,805 PYs. 
The red triangles correspond to the design parameters and the red lines correspond to settings when corresponding assumption holds (constancy assumption for (a) NI design and consistency of counterfactual placebo estimate for (b) AC-CF and (c) conservative AC-CF designs).}\label{fig:Violation_power}
\end{figure}

Figure \ref{fig:Violation_power} shows the empirical power for the three designs with different combinations of $(\lambda_P,\lambda_A)$.
All designs attain nominal power when the design parameters are correctly specified, and may attain sub-optimal power when the true parameters deviate from the design parameters, even when corresponding assumption holds.
For the NI design, the empirical power is smaller than the nominal level when the placebo incidence $\lambda_P$ is smaller than the design parameter.
For the AC-CF and conservative AC-CF designs, the empirical power is smaller than the nominal level when the incidence in the active control group $\lambda_A$ is smaller than the design parameter.

\subsection{Another setting with highly efficacious active control}
We consider another setting when the active control PrEP agent is expected to be highly efficacious.
Such a setting is likely in the future, for example, when CAB-LA is used as an active control.
We suppose that the active control is 90\% efficacious with $\log\lambda_P -\log\lambda_A = \log(10)$, such that $\lambda_A = 0.003$  cases/PY.
We will evaluate the RAE hypothesis (\ref{equ:RAE}) with the null hypothesis parameter $\gamma = 0.5$ under level $\alpha = 0.025$, and wish to achieve $\beta$-power against an alternative hypothesis with $\gamma^* = 1$ with $\beta = 0.8$ or $0.9$.
Under the null and alternative hypotheses, the prevention efficacy of the new PrEP agent is 68\% and 90\%, respectively, suggesting that the new PrEP agent is expected to be highly efficacious and we aim to rule out the possibility that the new PrEP agent is only moderately efficacious (i.e., the product is considered not sufficiently efficacious unless it reduces risk of infection by at least 68\%).
The remaining parameters are kept consistent with those outlined in Section \ref{sec:simu_para}.

Table \ref{tab:simu_CAB} shows the calculated sample sizes for the NI, AC-CF, and conservative AC-CF designs with a counterfactual placebo estimate based on external follow-up data.
In all designs, the empirical type-1 error rates are close to or fall below the nominal level, while empirical powers closely align with the nominal levels (data not shown). 
Figure \ref{fig:Violation_CABLA} in the Appendix shows the empirical type-1 error under violation of the assumptions for the three designs.
The NI and conservative AC-CF designs demonstrate a similar robustness against assumption violation as illustrated in Figure \ref{fig:Violation}.

When both the new PrEP agent and the active control are expected to be highly efficacious, the sample size for the NI design is substantially larger than that shown in Table \ref{tab:simu_alldesigns}.
For the AC-CF and conservative AC-CF designs, the calculated sample sizes are comparable or even lower than those shown in Tables \ref{tab:simu_alldesigns}.
In other words, despite potentially lower precision in the estimated incidences for both the new PrEP agent and the active control due to a smaller sample size in the trial, the power of the AC-CF and conservative AC-CF design remains attainable.
Therefore, the proposed AC-CF and conservative AC-CF are especially useful in the scenario of highly effective active control and new PrEP agent.

\begin{table}[htbp]
\small
\caption{Sample sizes for the NI, AC-CF and conservative AC-CF trial designs with a counterfactual placebo estimate based on external follow-up data when the active control is highly efficacious. 
For all designs, total PYs and total \#
Events include the prospective follow-up time and expected number of HIV infections under the alternative hypothesis, respectively, in the active-controlled trial only.}\label{tab:simu_CAB}
\vspace{.1in}
\begin{center}
\begin{tabular}{lccccc}
\hline
\multirow{2}{*}{Design}  & \multicolumn{2}{c}{Design power = 80\%} &&\multicolumn{2}{c}{Design power = 90\%}\\\cline{2-3}\cline{5-6}
& Total PYs & Total \#Events&& Total PYs & Total \#Events\\\hline
NI &	16738	&	50&&	22356	&	67\\
AC-CF - External follow-up &	5074	&	15&&	6858	&	21\\
Conservative AC-CF - External follow-up &	6378	&	19&&	8606	&	26\\
\hline
\end{tabular}
\end{center}
\normalsize
\end{table}

\subsection{Comparison with single-arm trial with a counterfactual placebo}\label{sec:simu_onearm}
As highlighted in the introduction, the incorporation of a randomized, active control arm in the prospective trial holds significant importance.
This inclusion facilitates a direct comparison of safety profiles and upholds the essential scientific integrity inherent in a randomized comparison.
Moreover, it may provide some robustness against bias of the counterfactual placebo estimate compared to the single-arm trial designs with a counterfactual placebo.

Nonetheless, in order to highlight the robustness that is afforded by including the active control arm in the AC-CF design, in this section, we compared the proposed designs to a single-arm trial with a counterfactual placebo.
This generalizes from the framework proposed by \cite{gao2021sample}.
The main hypothesis of interest is on the absolute efficacy of the new PrEP agent, given by
\[H_0^{E}: \log\lambda_P - \log\lambda_E\le \gamma_E \text{ vs. } H_a^{E}: \log\lambda_P - \log\lambda_E > \gamma_E,\]
where $\gamma_E$ suggests the acceptable efficacy of the new PrEP agent.
Based on data collected from the single-arm trial with a counterfactual placebo, the hypothesis can be evaluated through the test statistic
\[T^E = \frac{\log\hlambda_E^S - \log\hlambda_P - \gamma_E}{\hsigma_{PE}},\]
where $\hlambda_E^S$ is the estimated incidence for individuals that receive the new PrEP agent in the single-arm trial and $\hsigma_{PE}$ is the estimated standard error of the numerator.

In this simulation, we evaluate the performance of the single-arm design with a counterfactual placebo and compare it with the proposed AC-CF design with the same counterfactual placebo incidence estimate.
Both designs do not intentionally account for the bias of the counterfactual placebo estimate, and we evaluate how they perform under the same level of bias of the counterfactual placebo estimate.
For the single-arm design, we set $\gamma_E = \gamma (\log\lambda_P - \log\lambda_A) = 0.5\log(2.2) = 0.39$ and consider an alternative hypothesis with $\log\lambda_P - \log\lambda_E = \gamma_E^* = \gamma^* (\log\lambda_P - \log\lambda_A) = 1.36 \log(2.2) = 1.08$, such that the null and alternative hypotheses matches the RAE hypothesis in Section \ref{sec:simu_para}.
The rest of the design parameters are the same as in Section \ref{sec:simu_para} and \ref{sec:simu_violation}  and we target 80\% power.
The sample sizes for the AC-CF and single-arm designs are 4,942 and 2,398 trial PYs, respectively.

\begin{figure}[htbp] \centering
\includegraphics[width=\textwidth]{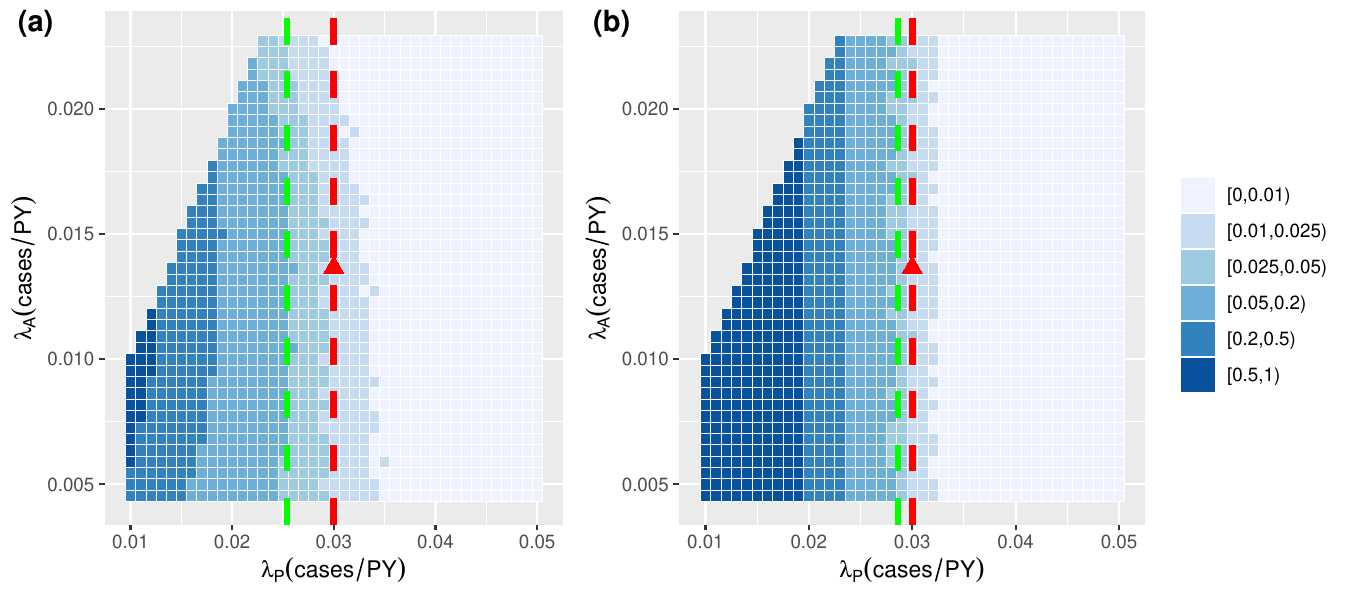}
\caption{Empirical type-1 error for assessing the hypothesis of interest with different values of $(\lambda_P,\lambda_A)$.
(a) AC-CF design to evaluate the RAE hypothesis with 4,942 trial PYs and a counterfactual placebo estimate based on external follow-up data with 1,805 PYs. (b) Single-arm design to evaluate the absolute efficacy hypothesis with 2,398 trial PYs and a counterfactual placebo estimate based on external follow-up data with 1,805 PYs.
The red triangles correspond to the design parameters.
The red lines correspond to settings when the counterfactual placebo estimate is consistent.
The green lines correspond to fitted lines with the empirical type-1 error close to 0.05 level.}\label{fig:onearm}
\end{figure}

Figure \ref{fig:onearm} shows the plot of the empirical type-1 error for the AC-CF and the single-arm designs in the scenarios when the counterfactual placebo estimate may not be consistent.
While both designs maintain protected type-1 error rates in cases where the counterfactual placebo estimate is consistent, it is notable that the inflation of type-1 error is more pronounced in the single-arm design with a counterfactual placebo, as illustrated by the green fitted lines denoting the scenarios that has a doubled type-1 error.
Therefore, the proposed AC-CF-type designs are more robust against bias of the counterfactual placebo estimate compared to the corresponding single-arm designs.

\section{Discussion}\label{sec:discuss}
This manuscript explored the design of an active-controlled trial incorporating a counterfactual placebo, along with proposed testing procedures and sample size calculations. 
Through simulation studies, we illustrated that the proposed design results in a reduced sample size compared to the NI design, indicating the initial feasibility of its implementation, particularly when the active control is highly efficacious. 
The conservative AC-CF design offers a conservative approach, given a natural concern about potential bias in the estimate of incidence using the counterfactual placebo. 
This aligns with the conservative approach used in the development of the NI design, where because of the reliance on the assumption that the active control was similarly effective to prior placebo-controlled trials, a cautious estimate of proven efficacy is used in setting the NI margin.
We suggest that, similarly, the conservative AC-CF design be applied in practice, rather than the AC-CF design. 

We also compare the proposed AC-CF designs with single-arm trials with a counterfactual placebo in our simulation studies.
Our findings demonstrate that while single-arm designs may require fewer participants, they are less robust against inconsistencies in counterfactual placebo estimates.
Additionally, these designs miss the opportunity to conduct a rigorous randomized comparison and gather comparative safety data for both the experimental intervention and the active control.

In our simulation studies, the primary comparison is between the proposed AC-CF-type designs and the NI design.
These two types of designs rely on different assumptions: the NI design relies on the constancy assumption, which posits that the efficacy of the active control remains the same in both historical and current trials.
On the other hand, the AC-CF-type designs depend on the consistency of counterfactual placebo incidence estimates.
Some may argue that the constancy assumption is more likely to hold.
However, in practice, the validity of each assumption depends on the specific population, the choice of active control, and reliability of the external data. 
Moreover, for a fair comparison, we assume a similar level of external information in the simulations.
In practical settings, the amount of statistical information available from the counterfactual placebo could be significantly larger or smaller.
Evaluating how the proposed designs perform in real-world settings is an important area for future research.

We propose a conservative AC-CF design that allows bias from counterfactual placebo incidence estimation and evaluate the acceptable level of bias through simulation studies.
In practice, conservativeness can also be incorporated by using more conservative estimates of placebo incidence, such as artificially increasing the variability of the estimate to account for model uncertainty.
Additionally, statistical methods for sensitivity analysis against unmeasured confounding \citep{imbens2015causal, fox2021applying} may be leveraged to formally account for the level of bias in the counterfactual placebo estimate in the proposed study designs.
These are important areas for future research.

We have assumed the availability of a counterfactual placebo incidence estimate and proposed designs that either assume its consistency or introduce additional protection against potential bias.
In practice, careful consideration is needed in choosing the approach for generating such a counterfactual placebo estimate and/or combining counterfactual placebo estimates from multiple approaches.
In addition to the commonly incorporated statistical estimation variability, potential for bias of the chosen estimation approach(es) and data need careful evaluation, for example,  the reliability of historical or concurrent data for generating the counterfactual placebo estimate.
 
We consider trial design based on testing the relative absolute efficacy hypothesis, motivated by its connection to the NI design.
A consideration is that the RAE is an unfamiliar scale for evaluating efficacy.
It is conceptually very similar to the averted infections ratio \citep{dunn2018averted, dunn2019connection, glidden2020bayesian}, with the latter based on incidence differences instead of incidence ratios.
The averted infections ratio offers potential advantages in terms of interpretability, and it may be more efficient particularly in scenarios with a limited number of events, as incidence difference often yields narrower confidence intervals. 
Moreover, other summary measures of the HIV infection endpoints, e.g., cumulative incidence, may also be of interest in defining prevention efficacy.
The proposed framework can be extended to accommodate those variations on prevention efficacy estimands.

We focused primarily on hypothesis testing that drives the sample size calculation, with particular attention to the performance of type-1 error when design assumptions are violated.
The power of the design is also crucial, and our simulation studies demonstrate that it depends not only on key assumptions (the constancy assumption for NI design and the consistency of the counterfactual placebo estimate for AC-CF-type designs) but also on other design parameters.
Incorporating conservativeness regarding power is another aspect of the design that requires further development.
Additionally, other design aspects, such as conducting interim monitoring, determining the analysis of endpoints at the study’s conclusion (i.e., efficacy estimation for the new PrEP agent), are of interest and need development.

Designing future HIV prevention trials with an active control with potentially 90\% prevention efficacy will be challenging.
The conservative AC-CF design we propose would require a trial size of 6,378 - 8,606 person years.
This is similar in size to successfully concluded recent evaluations of HIV vaccine, monoclonal and PrEP trials to date: e.g., 8,917 PYs in the HVTN 702 study \citep{gray2021vaccine}; 4,196 PYs in the HVTN 704/HPTN 083 study \citep{corey2021two}.
The design requires only a total of 19-26 observed events, in a context with a large contrast in incidence between the estimated placebo counterfactual and both trial arms.
This is in stark contrast to the NI trial design, which demands 50 - 67 events and 16,738 - 22,356 person years, trials 2 - 4 times the largest randomized trials conducted in HIV prevention to date.
It seems unlikely trials of this magnitude would be feasible.
In the regulatory approval process, trials with such a low number of events will require careful consideration of the statistical framework and assumptions, as we have enunciated in this work.
This framework will provide an approach for careful evaluation of the credibility of the evidence for efficacy.    

\bibliography{References}
\appendix

\section{Derivation on the RAE Hypothesis in the NI Design}\label{append:NI_deriv}
The NI trial is usually designed based on the desired type-1 error and power on the NI hypothesis (\ref{equ:NI}).
Since the test statistic $T_{NI}$ approximately follows the standard normal distribution under the null hypothesis $H_0^*: \log\lambda_E - \log\lambda_A = \delta$, the rejection region for an $\alpha$-level test is given by  $\{T_{NI}\le z_\alpha\}$.
Then, the sample size that gives rise to $1-\beta$ power against a specific alternative hypothesis $H_a^*:\log\lambda_E-\log\lambda_a = \delta^*$ satisfies $\sigma_{EA} = (\delta-\delta^*) /  (z_{1-\beta}-z_\alpha)$.

In this section, we would like to derive the type-1 error and power of the same test rejection $\{T_{NI}\le z_\alpha\}$ against the RAE hypothesis (\ref{equ:RAE}).
Consider the null hypothesis $K_0^*:(\log\lambda_P-\log\lambda_E)/(\log\lambda_P-\log\lambda_A) =  \gamma$ that is on the boundary of the null hypothesis space.
The type-1 error is given by
\begin{align*}
    &\Pr(T_{NI}\le z_\alpha | K_0^*) \\
    =& \Pr\left(\frac{\log\hlambda_E - \log\hlambda_A - \delta}{\hsigma_{EA}}\le z_\alpha \Bigg|K_0^*\right)\\
    =& \Pr\left(\frac{\sigma_{EA}Z_1 + \log\lambda_E - \log\lambda_A - (1-\gamma)\{\sigma_{PA0}Z_2 + \log\lambda_{P0}-\log\lambda_{A0} + z_\alpha \sigma_{PA0}\}}{\hsigma_{EA}}\le z_\alpha \Bigg|K_0^*\right)\\
    =& \Pr\left(\frac{\{\sigma_{EA}Z_1 -(1-\gamma)\sigma_{PA0}Z_2\}+ (1-\gamma)(\log\lambda_P - \log\lambda_A) - (1-\gamma)\{ \log\lambda_{P0}-\log\lambda_{A0} + z_\alpha \sigma_{PA0}\}}{\hsigma_{EA}}\le z_\alpha\right)\\
    =&\Pr\left(\frac{\sqrt{\sigma_{EA}^2 + (1-\gamma)^2\sigma_{PA0}^2}Z + (1-\gamma)\left\{\left(\log\lambda_P-\log\lambda_A\right)-\left(\log\lambda_{P0}-\log\lambda_{A0}\right)-z_\alpha\sigma_{PA0}\right\}}{\sigma_{EA}}\le z_\alpha\right)\\
    \consq & \Pr\left(\frac{\sqrt{\sigma_{EA}^2 + (1-\gamma)^2\sigma_{PA0}^2}Z - (1-\gamma)z_\alpha\sigma_{PA0}}{\sigma_{EA}}\le z_\alpha\right)\\
    =& E\left\{\Phi\left( z_\alpha\frac{\sigma_{EA}+(1-\gamma)\sigma_{PA0}}{\sqrt{\sigma_{EA}^2 + (1-\gamma)^2\sigma_{PA0}^2}}\right)\right\} \le \Phi(z_\alpha) = \alpha,
\end{align*}
where $Z_1$, $Z_2$ and $Z$ denote independent standard normal random variables, $\consq$ denotes the requirement of constancy assumption, and $\Phi(\cdot)$ denotes the cumulative distribution function of standard normal distribution.
The last expression has the expectation operator, since $\sigma_{EA}$ is a random variable due to its relationship with the random margin $\delta$.

That is, making use of a ``95\%-95\%'' margin $\delta$, the type-1 error is no greater than $\alpha$, such that it preserves type-1 error for assessing (\ref{equ:RAE}).
The value of the type-1 error depends on the relative size of the variance of log rate ratio in the trial $\sigma_{EA}^2$ and the variance of the product of the log rate ratio in the historical trial and the percentage reduction $(1-\gamma)^2\sigma_{PA0}^2$.
It tends to the nominal level when $\sigma_{EA}^2/\{(1-\gamma)\sigma_{PA0}\}^2$ tends to zero or tends to infinity, that is, the relative sizes of the active-controlled trial and the historical placebo-controlled trial are very different.
Simple derivation gives
\[\Pr\left(T_{NI}\le z_\alpha \Big| K_0^*, \frac{\sigma_{EA}^2}{(1-\gamma)^2\sigma_{PA0}^2} = x\right) = \Phi\left(z_\alpha \frac{1+x^2}{\sqrt{1+x}}\right).\]

\begin{figure}[htbp] \centering
\includegraphics[width=0.65\textwidth]{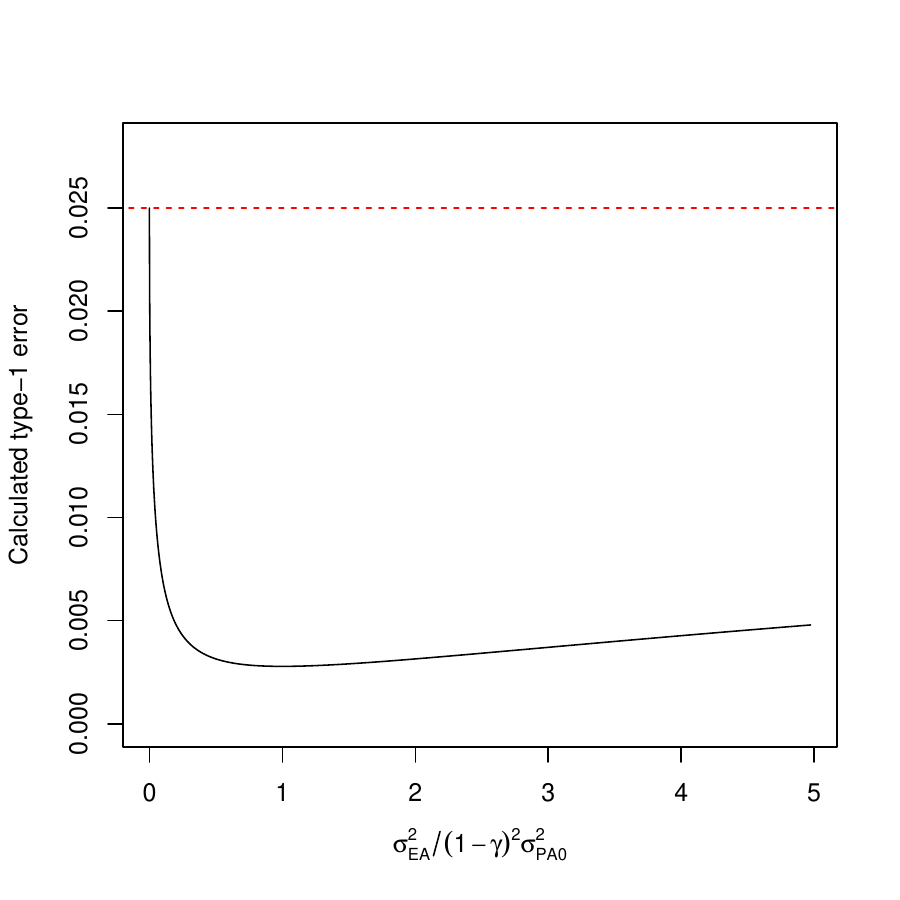}
\caption{ Calculated type-1 error for assessing the RAE hypothesis (\ref{equ:RAE}) with different values of $\sigma_{EA}^2/\{(1-\gamma)\sigma_{PA0}\}^2$. The type-1 error $\alpha = 0.025$.}\label{fig:NI_Ktype1}
\end{figure}

In practice, the type-1 error may be quite conservative numerically.
Figure \ref{fig:NI_Ktype1} shows the plot of the calculated type-1 error under $K_0^*$ when $\alpha = 0.025$ with different relative size of $\sigma_{EA}^2$ and $(1-\gamma)^2\sigma_{PA0}^2$.
The calculated type-1 error is far from the nominal value in a reasonable range of $\sigma_{EA}^2/\{(1-\gamma)\sigma_{PA0}\}^2$, e.g., in between 0.2 and 5, the calculated type-1 error is smaller than half of nominal level.
It reaches the smallest value (0.0028) when $\sigma_{EA}^2 = (1-\gamma)^2\sigma_{PA0}^2$.

To assess the power against the hypothesis on RAE (\ref{equ:RAE}), we define a specific alternative hypothesis 
\begin{equation}
    K_a^*: \frac{\log\lambda_P-\log\lambda_E}{\log\lambda_P-\log\lambda_A}= \gamma^*.\label{equ:RAE_alt_power}
\end{equation}
The power can be calculated as 
\begin{align*}
    \Pr(T_{NI}\le z_\alpha | K_a^*) =& E\left\{ \Pr\left(\frac{\log\hlambda_E - \log\hlambda_A - \delta}{\hsigma_{EA}}\le z_\alpha \Bigg|K_a^*,\delta\right)\right\}\\
    =&E\left\{\Pr\left(\frac{\sigma_{EA}Z + (1-\gamma^*)(\log\lambda_P-\log\lambda_A) - \delta}{\sigma_{EA}}\le z_\alpha\Bigg|\delta\right)\right\}\\
    =&E\left\{\Phi\left(z_{1-\beta} - \frac{(1-\gamma^*)(\log\lambda_P-\log\lambda_A)  - \delta^*}{\sigma_{EA}}\right)\right\},%\\
    %=&\Pr\left(Z +\frac{ (1-\gamma^*)(\log\lambda_P-\log\lambda_A) - \delta}{\delta-\delta^*}(z_{1-\beta}-z_\alpha)\le z_\alpha\right),
\end{align*}
where the last equality follows since the sample size requires $\sigma_{EA} = (\delta-\delta^*)/(z_{1-\beta}-z_\alpha)$.
The power is exactly $1-\beta$ when $\delta^* = (1-\gamma^*)(\log\lambda_P-\log\lambda_A)$, i.e., the alternative hypothesis for the NI hypothesis match with that for the relative absolute hypothesis.

\section{Technical Details for the AC-CF Design}\label{append:AC-CF}
Note that $K_0 = (H_a^{AS}\cap \tK_0)\cup (H_0^{AS}\cap \tK_a)$, such that for any specific null hypothesis $k_0\in K_0$, the type-1 error is given by
\begin{align*}
    &\Pr(T_{PA}\ge -z_\alpha, T_{CF}\ge -z_\alpha| k_0\in K_0)\\
    & \le \max \{\max_k \Pr(T_{PA}\ge -z_\alpha, T_{CF}\ge -z_\alpha| k\in H_a^{AS}\cap \tK_0), \max_k \Pr(T_{PA}\ge -z_\alpha, T_{CF}\ge -z_\alpha| k\in H_0^{AS}\cap \tK_a)\}\\
    & \le \max \{\max_k \Pr(T_{CF}\ge -z_\alpha| k\in H_a^{AS}\cap \tK_0),\max_k \Pr(T_{PA}\ge -z_\alpha| k\in H_0^{AS}\cap \tK_a)\}\\
    & = \max \{\max_k \Pr(T_{CF}\ge -z_\alpha| k\in \tK_0),\max_k \Pr(T_{PA}\ge -z_\alpha| k\in H_0^{AS})\}\\
    & = \max \{\Pr(T_{CF}\ge -z_\alpha| \tK_0^*),\Pr(T_{PA}\ge -z_\alpha| H_0^{AS,*})\} = \alpha,
\end{align*}
where $\tK_0^*: (1-\gamma)\log\lambda_P-\log\lambda_E+\gamma\log\lambda_A =0$ and 
$H_0^{AS,*}: \log\lambda_P-\log\lambda_A =0$.
That is, the type-1 error is always protected at $\alpha$-level
Similarly, under a specific setting with $(\lambda_P,\lambda_A)$, the power under a specific alternative hypothesis (\ref{equ:RAE_alt_power})
is given by
\begin{align*}
    \Pr(T_{PA}\ge -z_\alpha, T_{CF}\ge -z_\alpha| K_a^*,\lambda_A,\lambda_P) & \ge 1 - \Pr(T_{PA}< -z_\alpha| K_a^*,\lambda_A,\lambda_P)-\Pr(T_{CF}< -z_\alpha| K_a^*,\lambda_A,\lambda_P)\\   
    & = 1 - \Pr(T_{PA}< -z_\alpha| \lambda_A,\lambda_P)-\Pr(T_{CF}< -z_\alpha| K_a^*).  
\end{align*}
That is, if we a sample size $N$ would allow $\Pr(T_{CF} \ge -z_\alpha| K_a^*) - \Pr(T_{PA}< -z_\alpha | K_a^*) \ge \beta$, it will always guarantee a $\beta$-power against $K_a^*$ based on the proposed two-step testing procedure.

Note that we have 
\begin{align*}
    \Pr(T_{PA}< -z_\alpha | K_a^*,\lambda_P,\lambda_A) = & \Pr\left(\frac{\log\hlambda_P- \log\hlambda_A}{\hsigma_{PA}}<-z_\alpha\Big|\lambda_P,\lambda_A\right)
    =1-\Phi\left(z_\alpha + \frac{\log\lambda_P- \log\lambda_A}{\sqrt{(c_{P0} + c_A)/N + c_{P1}}}\right),
\end{align*}
and 
\begin{align*}
    \Pr(T_{CF}\ge -z_\alpha | K_a^*) =&\Pr\left(\frac{(1-\gamma)\log\hlambda_P - \log\hlambda_E + \gamma\log\hlambda_A}{\hV_\gamma^{1/2}}\ge -z_\alpha\Big|\frac{\log\lambda_P-\log\lambda_E}{\log\lambda_P-\log\lambda_A}=\gamma^*\right)\\
    =&\Phi\left(z_\alpha + \frac{(\gamma^*-\gamma)(\log\lambda_P - \log\lambda_A)}{\sqrt{\{(1-\gamma)^2c_{P0}+c_E+\gamma^2 c_A\}/N + (1-\gamma)^2c_{P1}}}\right).
\end{align*}
That is, the power against the alternative hypothesis $K_a^*$ also relies on the specific combination of $(\lambda_P,\lambda_A)$, since it impacts the distribution of $T_{PA}$.

To achieve $\beta$-power against the alternative hypothesis $K_a^*$, the sample size $N$ satisfies 
\begin{align*}
    &\Phi\left(z_\alpha + \frac{(\gamma^*-\gamma)(\log\lambda_P - \log\lambda_A)}{\sqrt{\{(1-\gamma)^2c_{P0}+c_E+\gamma^2 c_A\}/N + (1-\gamma)^2c_{P1}}}\right) + \Phi\left(z_\alpha + \frac{\log\lambda_P- \log\lambda_A}{\sqrt{(c_{P0} + c_A)/N + c_{P1}}}\right)= 1+\beta.
\end{align*}

\section{Technical Details for the Conservative AC-CF Design}\label{append:cons_AC-CF}
In this section, we calculate an analytical form for the type-1 error and power against the RAE hypothesis for the conservative AC-CF design.
Specifically, we demonstrate that the proposed conservative AC-CF design has protected type-1 error and nominal power against the RAE hypothesis.

\begin{figure}[htbp] \centering
\includegraphics[width=0.65\textwidth]{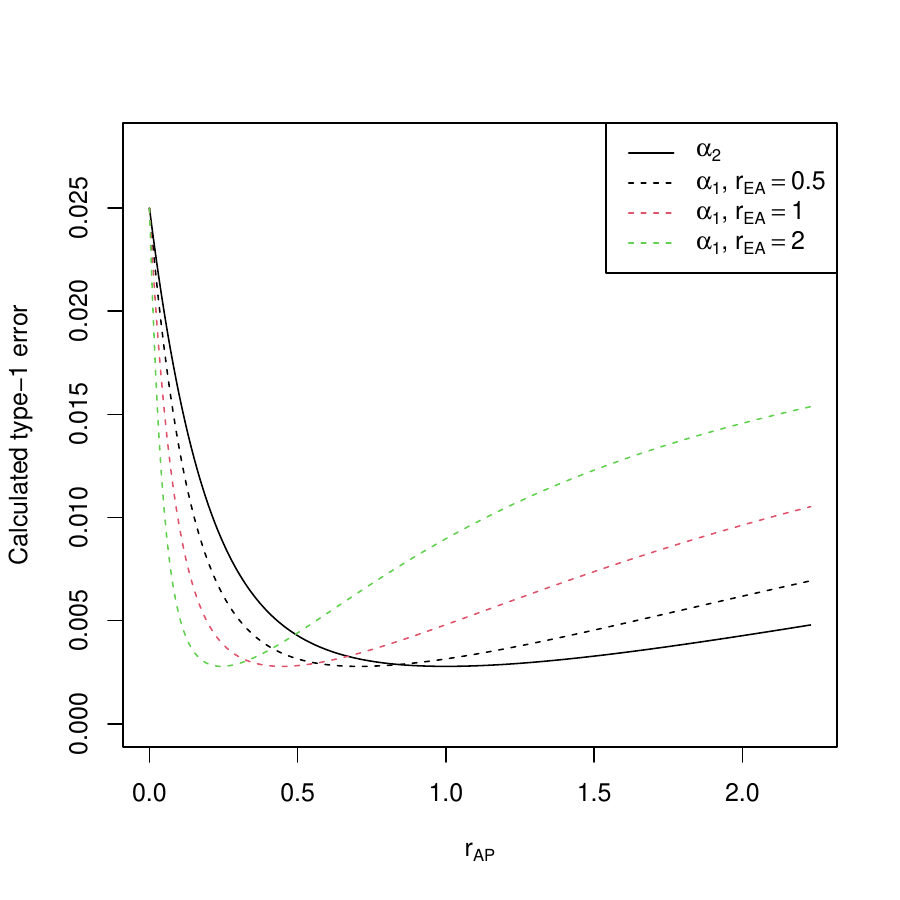}
\caption{ Calculated type-1 error in the conservative AC-CF design for assessing the RAE hypothesis (\ref{equ:RAE}) with different values of $r_{AP}$ and $r_{EA}$. The nominal type-1 error $\alpha = 0.025$.}\label{fig:cns_ACCF}
\end{figure}

Similar to the derivations in Appendix \ref{append:AC-CF}, the type-1 error of the conservative AC-CF design is bounded by $\max \{\Pr(T_{CF}'\ge -z_\alpha| \tK_0^*),\Pr(T_{PA}'\ge -z_\alpha| H_0^{AS,*})\}$.
Note that
\begin{align*}
    \Pr(T_{CF}'\ge -z_\alpha| \tK_0^*) =& \Pr\left(\frac{(1-\gamma)\log\hlambda_P^L - \log\hlambda_E + \gamma\log\hlambda_A}{\hV_\gamma'^{1/2}}\ge -z_\alpha \Bigg|\tK_0^*\right)\\
    =& \Pr\left(\frac{(1-\gamma)\{\log\lambda_P+ \sigma_PZ_1 + z_\alpha\hsigma_P\} - \log\lambda_E +\gamma \log\lambda_A + V_\gamma'^{1/2} Z_2}{\hV_\gamma'^{1/2}}\ge -z_\alpha \Bigg|K_0^*\right)\\
    =&\Pr\left(\frac{\sqrt{(1-\gamma)^2\sigma_P^2 + V_\gamma'}Z 
    + (1-\gamma)z_\alpha\sigma_P}{V_\gamma'^{1/2}}\ge -z_\alpha\right)\\
    =& \Phi\left( z_\alpha\frac{V_\gamma'^{1/2}+(1-\gamma)\sigma_P}{\sqrt{V_\gamma' + (1-\gamma)^2\sigma_P^2}}\right).
\end{align*}
Similarly, we have 
\begin{align*}
    \Pr(T_{PA}'\ge -z_\alpha| H_0^{AS,*})
    =& \Pr\left(\frac{\log\hlambda_P^L - \log\hlambda_A}{\hsigma_{A}}\ge -z_\alpha\Bigg|H_0^{AS,*}\right)\\
    =& \Pr\left(\frac{\log\lambda_P+ \sigma_PZ_1 + z_\alpha\hsigma_P - \log\lambda_A + \sigma_A Z_2}{\hsigma_A}\ge -z_\alpha \Bigg|H_0^{AS,*}\right)\\
    =&\Pr\left(\frac{\sqrt{\sigma_P^2 + \sigma_A^2}Z 
    + z_\alpha\sigma_P}{\sigma_A}\ge -z_\alpha\right)\\
    =& \Phi\left( z_\alpha\frac{\sigma_A+\sigma_P}{\sqrt{\sigma_A^2 + \sigma_P^2}}\right).
\end{align*}
Then the type-1 error of the conservative AC-CF design is bounded by
\[\max\left\{\Phi\left( z_\alpha\frac{V_\gamma'^{1/2}+(1-\gamma)\sigma_P}{\sqrt{V_\gamma' + (1-\gamma)^2\sigma_P^2}}\right),\Phi\left( z_\alpha\frac{\sigma_A+\sigma_P}{\sqrt{\sigma_A^2 + \sigma_P^2}}\right)\right\},\]
which depends on the relative size of $V_\gamma'$, $\sigma_A^2$, and $\sigma_P^2$.
In the special case when $\hlambda_P$, $\hlambda_A$, and $\hlambda_E$ are mutually independent, the type-1 error is given by $\max \{\alpha_1(r_{AP},r_{EA};\gamma),\alpha_2(r_{AP})\}$,
where $r_{AP} = \sigma_A/\sigma_P$, $r_{EA} = \sigma_E/\sigma_A$,
\[\alpha_1(r_{AP},r_{EA};\gamma) =  \Phi\left( z_\alpha\frac{\sqrt{r_{EA}^2+\gamma^2}r_{AP}+(1-\gamma)}{\sqrt{(r_{EA}^2+\gamma^2)r_{AP}^2 + (1-\gamma)^2}}\right)\text{, and } \alpha_2(r_{AP}) = \Phi\left( z_\alpha\frac{r_{AP}+1}{\sqrt{r_{AP}^2 + 1}}\right).\]
We plot in Figure \ref{fig:cns_ACCF} the values of $\alpha_1$ and $\alpha_2$ with different values of the ratios $r_{AP}$ and $r_{EA}$ when $\alpha = 0.025$ and $\gamma = 0.5$.
The type-1 error, given by the maximum of $\alpha_1$ and $\alpha_2$, is always preserved.
It is obtained by $\alpha_1$ when $r_{AP}$ is large and is obtained by $\alpha_2$ when $r_{AP}$ is small.
It increase as $r_{EA}$ increases and when $r_{AP}$ is large.

Similarly, the power under a specific alternative hypothesis (\ref{equ:RAE_alt_power})
is given by
\begin{align*}
    \Pr(T_{PA}'\ge -z_\alpha, T_{CF}'\ge -z_\alpha| K_a^*,\lambda_A,\lambda_P) & \ge 1 - \Pr(T_{PA}'< -z_\alpha| \lambda_A,\lambda_P)-\Pr(T_{CF}'< -z_\alpha| K_a^*)\\
    & = \Pr(T_{PA}'\ge -z_\alpha| \lambda_A,\lambda_P)+\Pr(T_{CF}'\ge  -z_\alpha| K_a^*)-1.
\end{align*}
Note that
\begin{align*}
    \Pr(T_{PA}'\ge -z_\alpha| \lambda_P,\lambda_A)
    =& \Pr\left(\frac{\log\hlambda_P^L - \log\hlambda_A}{\hsigma_{A}}\ge -z_\alpha\Bigg|\lambda_P,\lambda_A\right)\\
    =& \Pr\left(\frac{\log\lambda_P+ \sigma_PZ_1 + z_\alpha\hsigma_P - \log\lambda_A + \sigma_A Z_2}{\hsigma_A}\ge -z_\alpha \Bigg|\lambda_P,\lambda_A\right)\\
    =&\Pr\left(\frac{(\log\lambda_P-\log\lambda_A)+\sqrt{\sigma_P^2 + \sigma_A^2}Z 
    + z_\alpha\sigma_P}{\sigma_A}\ge -z_\alpha\right)\\
    =& \Phi\left( \frac{\log\lambda_P-\log\lambda_A}{\sqrt{\sigma_A^2 + \sigma_P^2}}+ z_\alpha\frac{\sigma_A+\sigma_P}{\sqrt{\sigma_A^2 + \sigma_P^2}}\right),
\end{align*}
and 
\begin{align*}
    \Pr(T_{CF}'\ge -z_\alpha| K_a^*)
    =& \Pr\left(\frac{(1-\gamma)\{\log\lambda_P+ \sigma_PZ_1 + z_\alpha\hsigma_P\} - \log\lambda_E +\gamma \log\lambda_A + V_\gamma'^{1/2} Z_2}{\hV_\gamma'^{1/2}}\ge -z_\alpha \Bigg|K_a^*\right)\\
    =&\Pr\left(\frac{(\gamma^*-\gamma)(\log\lambda_P - \log\lambda_A)+\sqrt{(1-\gamma)^2\sigma_P^2 + V_\gamma'}Z 
    + (1-\gamma)z_\alpha\sigma_P}{V_\gamma'^{1/2}}\ge -z_\alpha\right)\\
    =& \Phi\left( \frac{(\gamma^*-\gamma)(\log\lambda_P - \log\lambda_A)}{\sqrt{V_\gamma' + (1-\gamma)^2\sigma_P^2}} + z_\alpha\frac{V_\gamma'^{1/2}+(1-\gamma)\sigma_P}{\sqrt{V_\gamma' + (1-\gamma)^2\sigma_P^2}}\right).
\end{align*}
That is, if the sample size $N_{95\%-95\%}$ satisfies
\begin{align*}
    &\Phi\left( \frac{\log\lambda_P-\log\lambda_A}{\sqrt{\sigma_A^2 + \sigma_P^2}}+ z_\alpha\frac{\sigma_A+\sigma_P}{\sqrt{\sigma_A^2 + \sigma_P^2}}\right) + \Phi\left( \frac{(\gamma^*-\gamma)(\log\lambda_P - \log\lambda_A)}{\sqrt{V_\gamma' + (1-\gamma)^2\sigma_P^2}} + z_\alpha\frac{V_\gamma'^{1/2}+(1-\gamma)\sigma_P}{\sqrt{V_\gamma' + (1-\gamma)^2\sigma_P^2}}\right) =1+\beta,
\end{align*}
then a $\beta$-power is guaranteed.

\begin{figure}[htbp] \centering
\includegraphics[width=\textwidth]{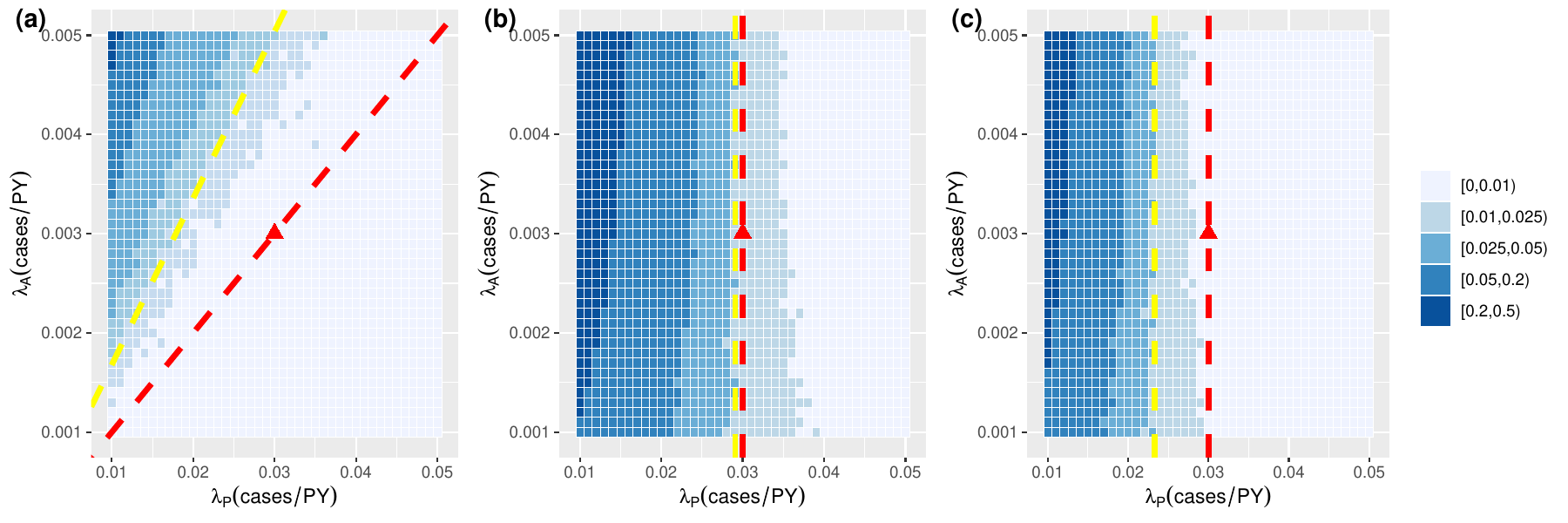}
\caption{Empirical type-1 error for assessing the RAE hypothesis with different values of $(\lambda_P,\lambda_A)$ when the new PrEP agent and active control are expected to be highly efficacious.
(a) NI design with an average of 16,738 PYs and 95\%-95\% margin based on a historical placebo-controlled trial with 3,610 PYs. (b) AC-CF design with 5,074 trial PYs and a counterfactual placebo estimate based on external follow-up data with 1,805 PYs. (c) Conservative AC-CF design with 6,378trial PYs and a counterfactual placebo estimate based on external follow-up data with 1,805 PYs. 
The red triangles correspond to the design parameters.
The red lines correspond to settings when corresponding assumption holds (constancy assumption for (a) NI design and consistency of counterfactual placebo estimate for (b) AC-CF and (c) conservative AC-CF designs).
The yellow lines correspond to fitted lines with the empirical type-1 error close to the nominal 0.025 level.}\label{fig:Violation_CABLA}
\end{figure}

\begin{table}[htbp]
\small
\caption{Screening sample sizes in the AC-CF and conservative AC-CF trial designs with a counterfactual placebo based on recency testing at screening.}\label{tab:simu_RA}
\vspace{.1in}
\begin{center}
\begin{tabular}{ccccccccc}
\hline
Design	&	Follow-up	&	\multicolumn{3}{c}{AC-CF Design}					&&	\multicolumn{3}{c}{Conservative AC-CF Design}	\\\cline{3-5}\cline{7-9}
Power	&	Year	&	\#Screened	&	\#HIV+	&	\#Recent	&&	\#Screened	&	\#HIV+	&	\#Recent	\\\hline
0.8	&	1	&	6391	&	959	&	70	&&	9725	&	1459	&	106	\\
0.8	&	2	&	3922	&	588	&	43	&&	6158	&	924	&	67	\\
0.9	&	1	&	8080	&	1212	&	88	&&	11920	&	1788	&	130	\\
0.9	&	2	&	4939	&	741	&	54	&&	7518	&	1128	&	82 \\
\hline
\end{tabular}
\end{center}
\normalsize
\end{table}

\end{document}